\documentclass[twocolumn,showpacs,preprintnumbers]{revtex4}
\usepackage{mathrsfs}
\usepackage{amssymb}
\usepackage{dsfont}
\usepackage{graphicx}
\usepackage{bm}

\graphicspath{%
    {converted_graphics/}
    {/}
}
\begin{document}

\title{Experimental evidence of twin fast metastable  H($2^2$S) atoms from dissociation of  cold H$_2$ induced by electrons}
\author{J. Robert$^{1}$, F. Zappa$^{2}$, C.R. Carvalho$^{3}$, Ginette Jalbert$^{3}$, R.F.Nascimento$^{4}$, A. Trimeche$^{1}$, O. Dulieu$^{1}$, Aline Medina$^{5}$, Carla Carvalho$^{6}$ and N. V. de Castro Faria$^{3}$
\\
$^1$ Laboratoire Aim\'{e} Cotton, CNRS/Univ Paris-Sud/ENS Cachan, B\^at.505, Campus d'Orsay, 91405 Orsay Cedex, France 
\\
$^2$Departamento de F\'\i sica, UFJF, Minas Gerais 36036-330, Brazil
\\
$^3$Instituto de F\'\i sica, UFRJ, Cx.~Postal 68528, Rio de Janeiro 
21941-972, Brazil
\\
$^4$Centro Federal de Educa\c c\~{a}o Tecnol\'{o}gica Celso Suckow da Fonseca, 25620-003, Petr\'{o}polis, RJ, Brazil
\\
$^5$Instituto de F\'\i sica, Universidade Federal da Bahia, Salvador, BA 40210-340, Brazil
\\
$^6$ Instituto de Aplica\c c\~ao Fernando Rodrigues da Silveira, UERJ 20261-232, Rio de 
Janeiro, RJ, Brazil}

\date{\today}

\begin{abstract}
We report the direct detection of two metastable H($2^2$S) atoms coming from
the dissociation of a single cold H$_2$ molecule, in coincidence measurements. The molecular dissociation was induced by electron impact in order to avoid limitations by the selection rules governing radiative transitions. Two detectors, placed close from the collision center, measure the neutral metastable H($2^2$S) through a localized quenching process, which mixes the H($2^2$S) state with the H($2^2$P), leading to a Lyman-$\alpha $ detection. Our data show the accomplishment of a coincidence measurement which proves for the first time the existence of the H($2^2$S)-H($2^2$S) dissociation channel.
\end{abstract}


\maketitle

In the 60s, M. Leventhal, R. T. Robiscoe and K. R. Lea \cite{Leventhal67} performed the first measurements of the time-of-flight (TOF) and angular distribution of H($2^2$S) atoms produced by H$_2$ dissociation. Since then the structure of H$_{2}$ and the dynamics of its fragmentation channels are subjects of renewed experimental studies, by synchrotron radiation or laser techniques or in ion beam experiment \cite{alkhalili04,wolf05,zhaunerchyk07} via the study of dissociative recombination \cite{zande09,larsson-orel08}, as well as theoretical ones, by {\it ab initio} calculations or by the method of the molecular multichannel quantum defect \cite{Worner07,Melero06,Bozek06,Sanchez97.99,Vanne06,Martin07,mot08}. Despite the wealth of information on fragmentation of H$_{2}$, according to our knowledge, the production of two metastable atoms H($2^2$S) arriving from the dissociation of the same molecule  has not been reported yet.

The fact that until now no research group has reported the evidence of detection of twin H($2^2$S)-H($2^2$S) atoms should be a consequence of the small cross section related to this dissociation channel. Consequently, the corresponding coincidence rate will be small and care has to be taken as well as specific experimental conditions have to be fulfilled to observe it.

We have experienced this difficulty in the last years trying to observe this channel.  In the beginning, our setup was an adaptation of a high-precision {\it time-of-flight} spectroscopy experiment, with which we have studied the production of fast and slow metastable H($2^2$S) atoms coming from the dissociation of cold H$_{2}$ molecules \cite{Medina11,Medina12}. Since then we have believed we were able to estimate, according to our collision kinematics analysis, the precise direction in which the twin atoms should be found and, consequently, we used a detection system that is capable of positioning the detectors´s active area with very high precision, providing an extremely narrow solid angle associated to the region, from which we detected the H($2^2$S) atoms. For details we address the reader to our previous works mentioned earlier. The joint effect of that small solid angle together with the small cross section of the H($2^2$S)-H($2^2$S) dissociation channel resulted in that we were able only to observe the random coincidence rate of H($2^2$S) atoms from the H($2^2$S)-H($2^2$P) dissociation channel.  According to that, we mounted a second experiment. The difference was this new setup was not an adaptation from a previous one, but it was designed to our specific purpose. This second experiment involved a small collision chamber, which forced us to place the detectors directed to the collision region and close to it; besides, we did not have space to use the detection system that surrounds the detector as in the first experiment. The consequence of that was we lost direction precision, concerning the place from where we detect the H($2^2$S), but we gained with a very much larger solid angle which increased enormously the coincidence rate. Within this condition we have started to see coincidence rate which could be assigned to the H($2^2$P)-H($2^2$P) dissociation channel and another one which we recognized as being the twin H($2^2$S)-H($2^2$S). Then we returned to the first setup, adjusting it according to the conditions of the second setup in order to obtain the results we display in this letter.

Thus, in the present letter we discuss our experimental setup and the analysis of the data that provide evidence of the coincidence measurement of two H($2^2$S) atoms arriving from the same H$_2$ molecule. The experimental setup is in general the same which was discussed in recent papers of our group \cite{Medina11,Medina12}. In the next we briefly remind the experimental detail and then we discuss our data.

Our apparatus consists of a supersonic jet source of molecular hydrogen, whose stream crosses an electron beam coming from a pulsed electron gun. In all results appearing in this paper we have used a 100 ns electron pulse. The detection of the metastable atoms is performed by two detectors directed to the collision zone, placed symmetrically with respect to the plane defined by the electron and the H$_2$ beams. In Ref.\cite{Medina11,Medina12} we were concerned with the precision of the time-of-flight spectra, which led us to use a specially designed detection system, which ensured we read only H($2^2$S) atoms, and placed them as far as possible from the collision region. In the present case we have been concerned in enhancing the coincidence signals. Thus, we brought them closer, placing them at 58 mm and 67 mm from the collision region and we have used detectors facing the collision region, where the channeltron cone is protected by a grounded grid. We have maintained the detectors at different distances from the collision region, in order to avoid that the coincident counts overlap with any electric noise which could be picked up simultaneously by both detectors. Since there is direct line-of-sight between the active area of the detectors and the interaction region, we can detect, besides excited atoms, both UV radiation and scattered electrons coming from there. Ions are excluded since there is no acceleration for them to gain energy and efficiently produce secondary electrons at the front surface of the detectors. The time-of-flight system enable us to separate the Lyman Alpha radiation emitted by H($2^2$S) from all the other contributions. The signals from the detectors are separately pre-amplified and amplified by standard NIM electronics and a commercial time analyser card (FastComTec 7888) process and records the data. Throughout the experiment we have set the time analyser card with a time resolution of 16 ns per channel ({\it bin}).

\begin{figure}[h] 
  \centering
  \includegraphics[bb=119 49 476 793,width=5cm,height=12.5cm,keepaspectratio]{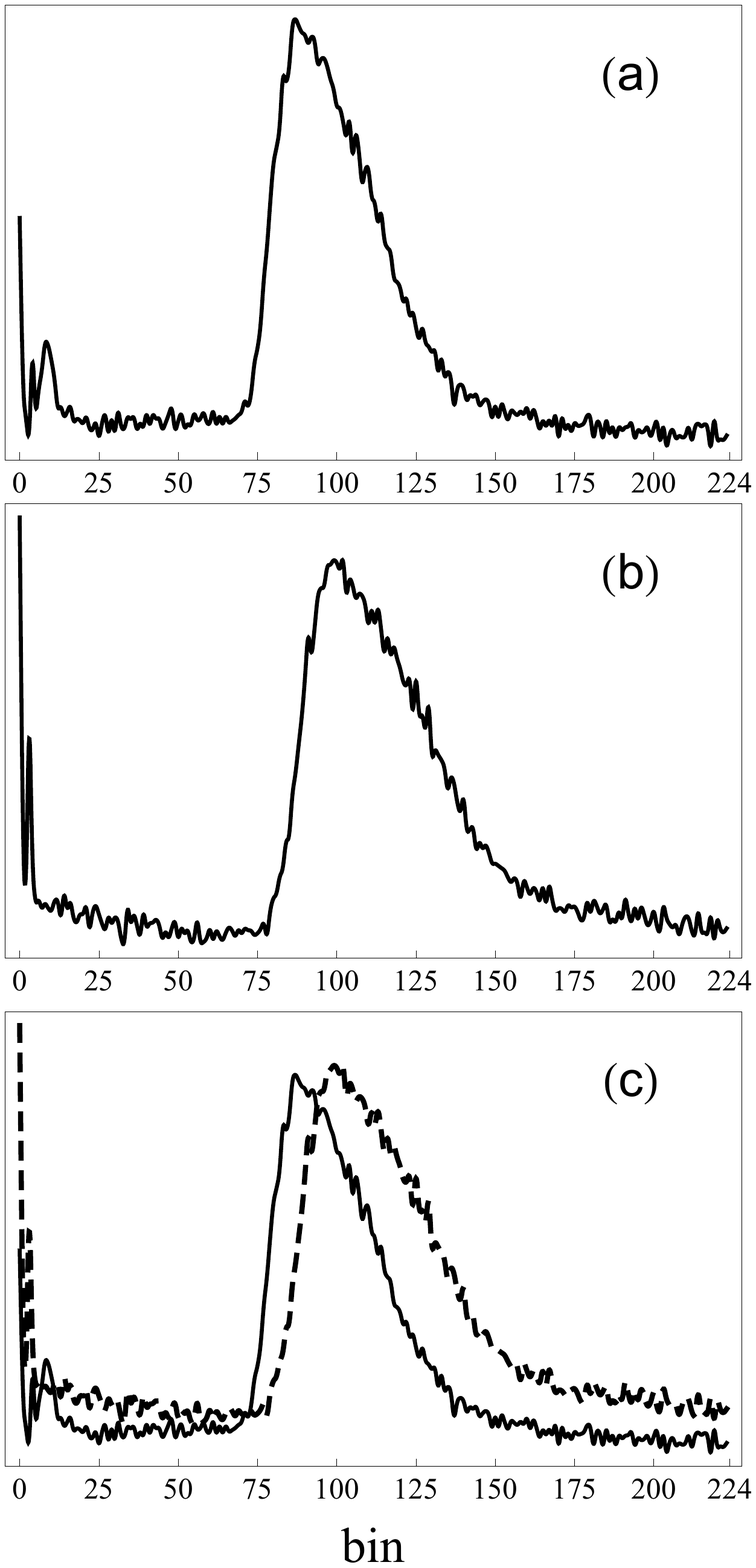}
  \caption{Time of flight spectrum obtained with detectors A and B placed opposite to each other with respect to the collision plane and at distance of (a) 58 mm and (b) 67 mm from the H$_{2}$ dissociation spot, respectively. The H$_{2}$ excitation is accomplished by electron impact. The energy of the electrons is 200 eV and the electron pulse width is 100 ns. The horizontal axis displays time in units ({\it bins}) of 16 ns. The vertical axis corresponds to couting rate in arbitrary units.(c) Spectra A (full line) and B (dashed line) together displaying the shift associated to the distance difference.}
  \label{Fig1}
\end{figure}

In Figs.\ref{Fig1}a and \ref{Fig1}b we show the time-of-flight (TOF) spectrum registered by both detectors, A and B, respectively. On the very left of both graphs we can observe a vertical line which corresponds to the H($2^2$P) atoms, whose lifetime is of the order of one ns, and to the photons produced by the excitation of the radiative states of the hydrogen molecule. The contributions that can be attributed to scattered electron or molecular hydrogen ions or ions produced by the excitation process appear next to the photon peak and can be more clearly seen in Fig.\ref{Fig1}a. We observe that its width is of the order of 100 ns as expected. In Fig.\ref{Fig1}b we also observe the presence of the scattered electrons, although not so clear. Finally in both figures we see the (large) peak corresponding to the {\it fast} H($2^2$S) atoms. The peak of the {\it slow} atoms produced in the H$_{2}$ dissociation does not appear there and it is further to the right. In Fig.\ref{Fig1}c we display both spectra in the same diagram to make clear the shift in time between them due to the different position of the detectors.

Now, let us discuss our coincidence measurement. The {\it start} of each sweep is established by the beginning of the electron beam pulse and the signal coming from each detector is fed into a different {\it stop} input of our time-of-flight acquisition card. Therefore, given a specific {\it start} signal, pulses coming from one detector are labeled as {\it stopA} and the other detector as {\it stopB}. Notice that a {\it start} not necessarily has a {\it stopA} and  a {\it stopB}. Given our relatively low count rate, most {\it start} pulses are not accompanied by a corresponding {\it stop} on either detector. The analyser card records all events where there was at least one {\it stopA} for a given {\it start}, and the corresponding{\it stopB} signals if they ocurred as well.

Therefore, our data would be a collection of pairs $(\tau_A,\tau_B)$, where $\tau_A (\tau_B)$ is the time interval elapsed between a {\it start} and its corresponding {\it stopA} ({\it stopB}). In fact our data are expressed in terms of pairs of $bins$ $(\beta_A, \beta_B)$. The FastComTec card gives the information directly in ns, in the following way: when we say that a particle arrives within a $bin$ $\beta$, we mean during the time window given by [$\tau, \tau + \Delta t$], where $\tau=\beta(\Delta t+\delta t)$, $\Delta t$ is the $bin$'s width (16 ns in the present case) and $\delta t$ is the $deadtime$ between two consecutive $bins$ (an experimental limitation of our analyser card; $\delta t= 0,5$ ns). In addition, the analyser card provide us the number of time $N_{AB}(\beta_A, \beta_B)$ that the pair $(\beta_A, \beta_B)$ appears, which yields the {\it coincidence counting spectrum} as a two-dimensional counting histogram $N_{AB}(\beta_A, \beta_B) \times \beta_A \times \beta_B$.

\begin{figure}[htbp] 
  \centering
  \includegraphics[bb=63 208 579 712,width=7cm,height=6.84cm,keepaspectratio]{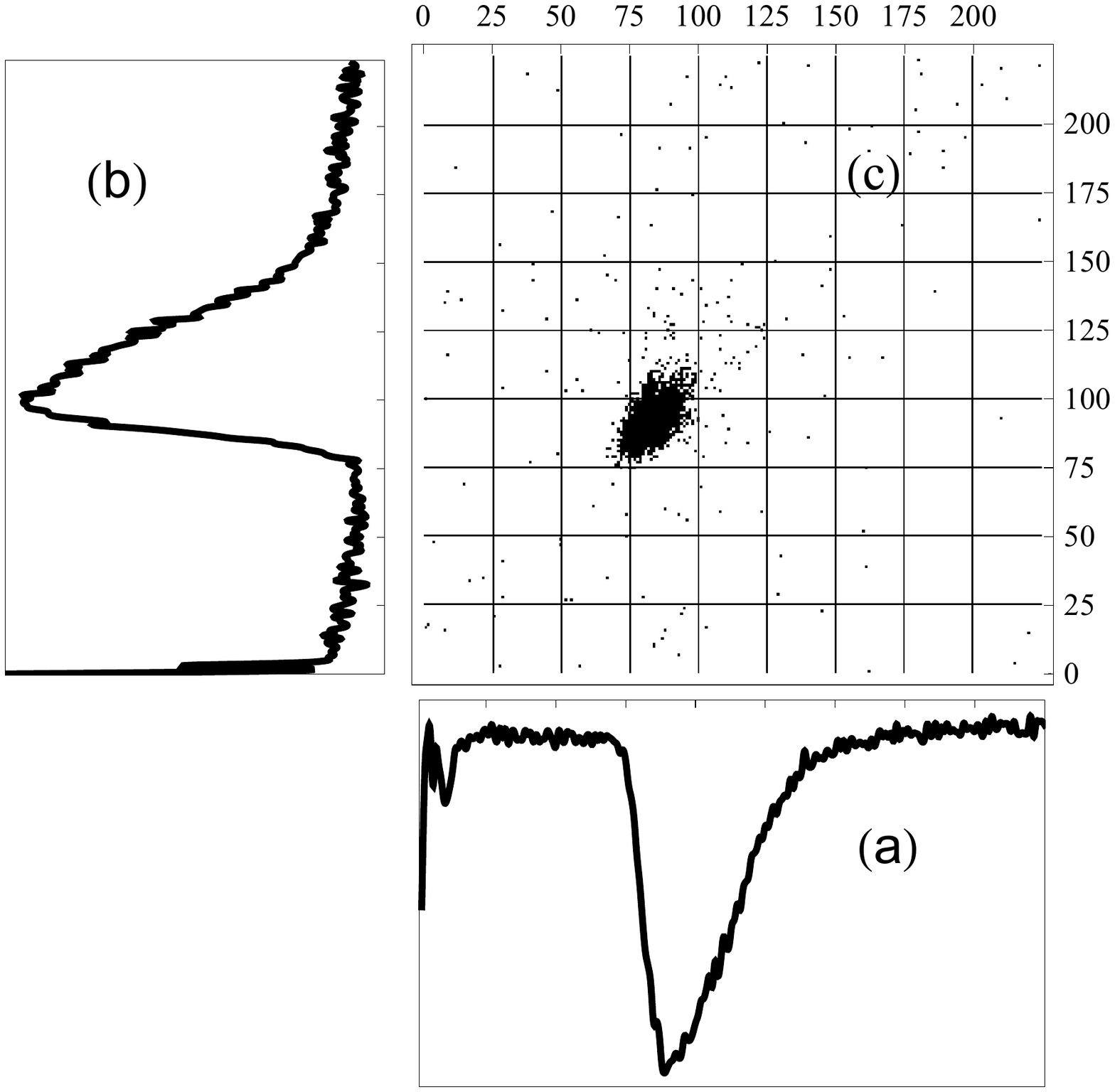}
  \caption{(a) Same as Fig.1a. (b)Same as Fig.1b. (c) Coincidence counting spectrum, for the same experimental conditions of Fig.1, containing a total of 3164 coincidence counts.}
  \label{Fig2}
\end{figure}

\begin{figure}[h] 
  \centering
  \includegraphics[bb=2 132 593 711,width=8cm,height=8cm]{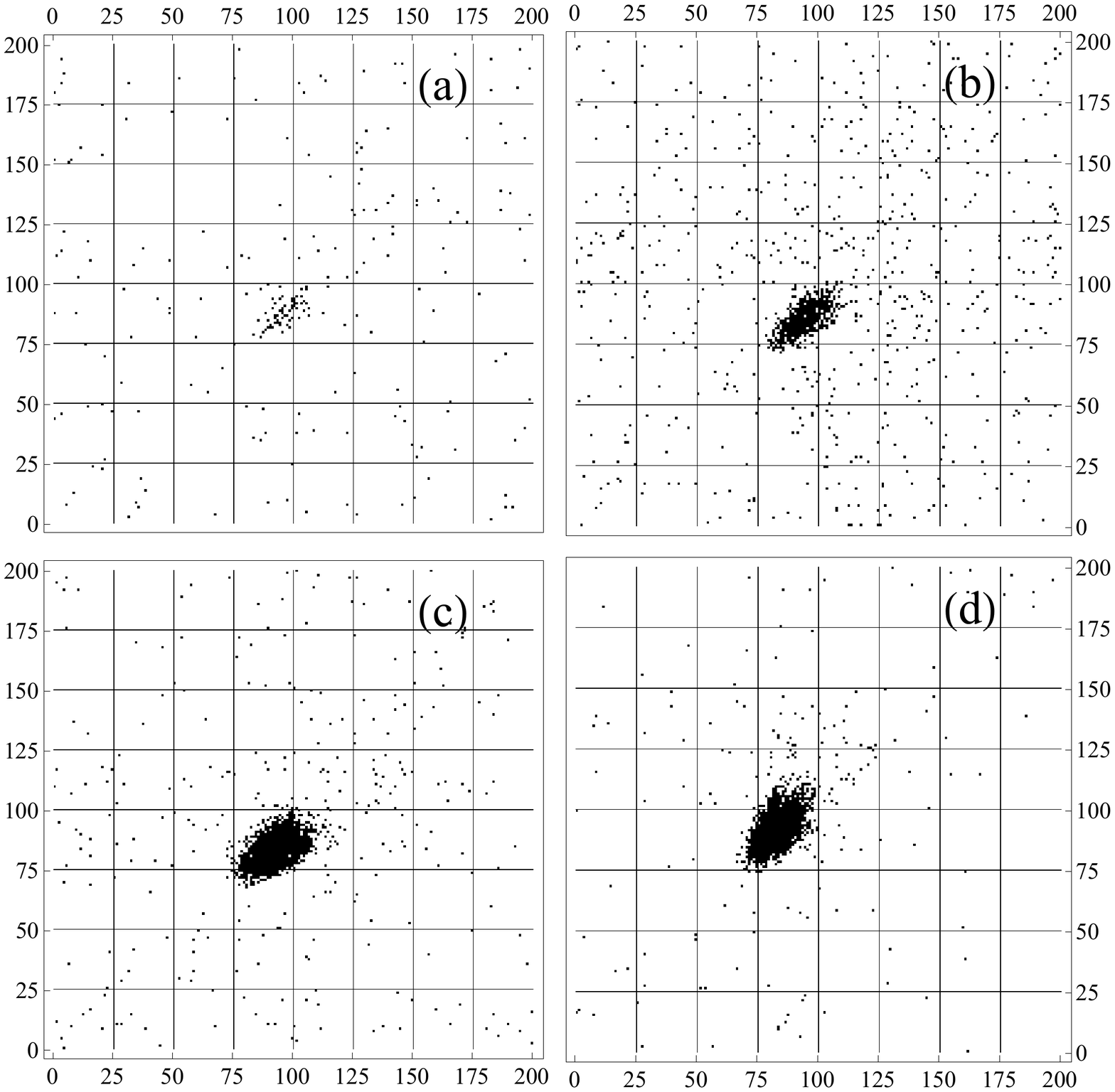}
  \caption{Top vision of the coincidence counting rate spectrum for different electrons energies (a) 60 eV (b) 80 eV (c) 200 eV (d) 200 eV.  In  (a) to (c) $L_A$= 67 mm and $L_B$=58 mm and in (d) $L_A$= 58 mm and $L_B$=67 mm.}
  \label{Fig3}
\end{figure}

In Fig.\ref{Fig2} we display the top vision of the coincidence counting spectrum (Fig.2c) with its corresponding TOF spectra placed accordingly to the axis associated to each one; on the bottom panel (Figs.2a) we have the former Fig.1a and, on the left panel, Fig.1b. The coincidence spectrum reveals random coincidences occurring over the whole domain in the $\beta_A \times \beta_B$ plane with a very low count. It also shows a concentration of coincidences in a region which corresponds to the first third part of the {\it fast} H($2^2$S) peak on both detectors. Besides, the coincidences which occur in that region have a counting rate much higher than average, as we can see in Fig.4 below, forming a peak inside the region expected to find the pair H($2^2$S)-H($2^2$S). This result was obtained during a round of the experiment which lasted 2h8min and yielded a total of 3164 coincidence counts.

In Fig.\ref{Fig3} we display the top vision of the coincidence counting rate spectrum for different electrons energies.    Besides, regarding to Fig.\ref{Fig1} and Fig.\ref{Fig2} the distance of the detectors were interchanged. Fig.\ref{Fig3}a shows the result obtained when the electrons energy was 60 eV, with a data acquisition period of 5h51min and resulted in a total of 286 coincidence counts. Fig.\ref{Fig3}b shows the result taken during 1h56min, for electrons with energy of 80 eV, involving a total counting number of 1424. Observe that, in spite of the lower acquisition time, the total counting number increased considerably in the whole $\beta_A \times \beta_B$ space. Besides, the small concentration, which barely appeared in Fig.\ref{Fig3}a, became stronger revealing now the coincidence peak. In Fig.\ref{Fig3}c it is shown the result for electrons with energy of 200 eV electrons and a period of 9h17min of data acquisition. The total counting number in this case is 3362. As a result, when normalised with respect to the electron flux the relative cross section for pair production is in arbitrary units 1 for 60 eV, 1227 for 80 eV and 42 for 200 eV. This shows that the process obeys to a threshold law. 

\begin{figure}[htbp] 
  \centering
  \includegraphics[bb=12 13 700 425,width=10cm,height=5.46cm,keepaspectratio]{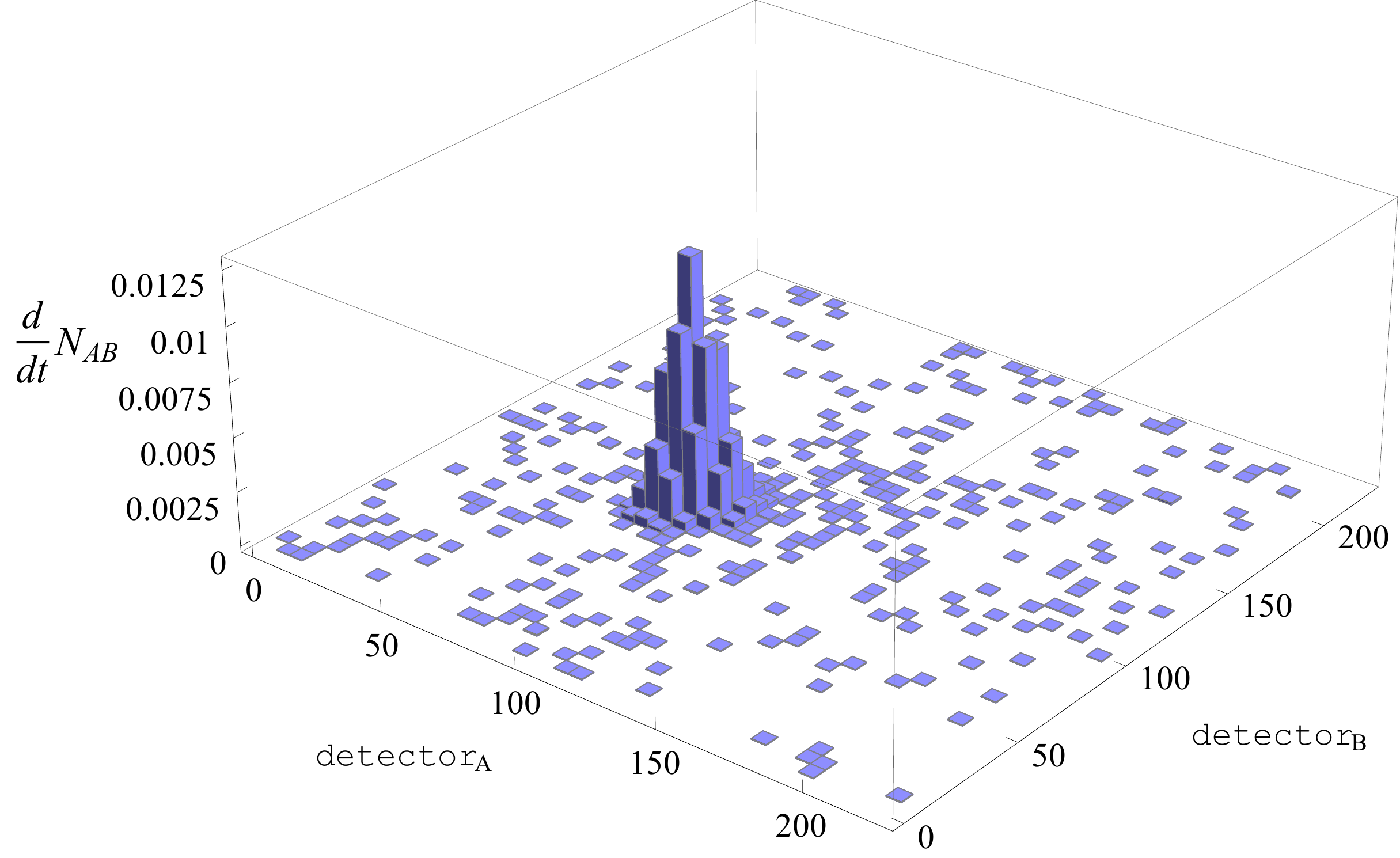}
  \caption{Coincidence counting rate spectrum as function of the bins   for the electron energy 200 eV and $L_A$= 67 mm and $L_B$=58 mm}
  \label{Fig4}
\end{figure}

\begin{figure}[htbp] 
  \centering
  \includegraphics[bb=12 13 700 425,width=10cm,height=5.45cm,keepaspectratio]{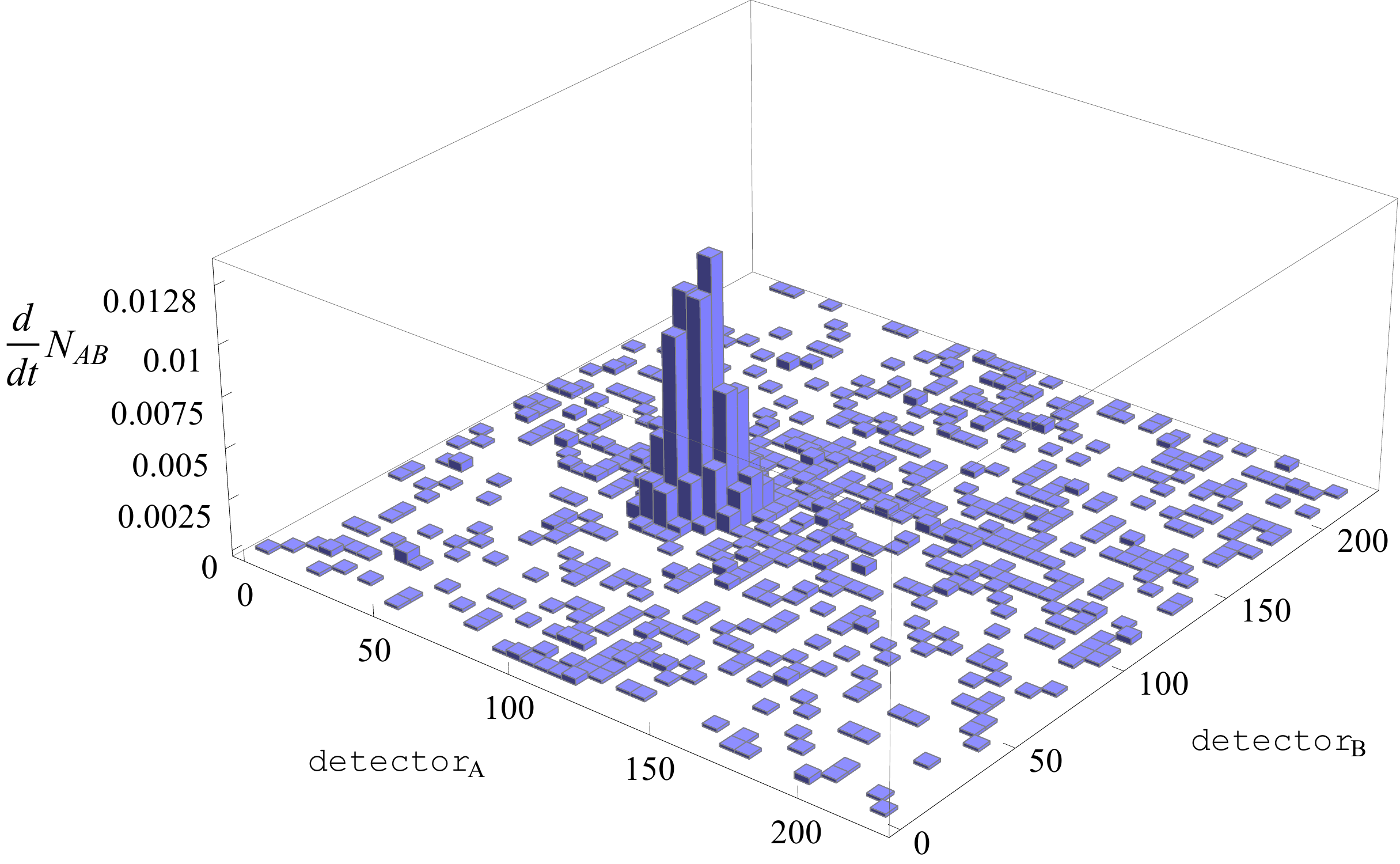}
  \caption{Coincidence counting rate spectrum as function of the bins for the electron energy  of 80 eV and $L_A$= 67 mm and $L_B$=58 mm}
  \label{Fig5}
\end{figure}

In Figs.\ref{Fig4} and \ref{Fig5}, we display a three dimensional  representation of the coincidence counting rate spectrum for different electrons energies corresponding to Fig.\ref{Fig3}c (200 eV) and Fig.\ref{Fig3}b (80 eV). Comparison between the two 3D spectrum  show clearly that the  number of random counts rate to the true coincidences is lower for the higher energy.

The results that we have described in this article are of multiple interests: first, they may guide theoretical computations regarding the doubly excited states of molecular hydrogen which asymptotically yield a pair H($2^2$S)-H($2^2$S), still an unsolved problem; second, an emerging pair of atoms with mean life time of the order of one tenth of second and  with polarized angular momentum (polarized spin or  polarized total angular momentum) may provide a new manner to obtain insight into the complex field of the molecular interactions, from the short-distance to the long-distance domain of interactions between moving atoms\cite{Beguin}. In a next stage, we will try to  follow the original ideas of Bohm \cite{Bohm}, when he proposed a variant of the EPR system \cite{Einstein}, involving a Stern-Gerlach interferometer.  In fact, a twin-atom experiment should be closer to the experimental suggestions of Englert, Schwinger and Scully, who followed Bohm's ideas (see for instance \cite{Schwinger}).

It is worth to mention that other proposals of experimental realizations of Bohm's spin-1/2 particle experiment, in its original form by making use only of atomic fragments and not photons, were already suggested \cite{Fry}. However, according to our knowledge, the production of two atoms arriving from the dissociation of the same molecule with the possibility of direct manipulation of them, according to Bohm's ideas, has not been achieved yet. On the other hand, recently another group has performed measurements involving the photodissociation of H$_2$ and has analysed the entanglement of the H atoms generated in this process by detecting the two Lyman-$\alpha$ photons emitted by the pair of short-life H($2^2$P) atoms \cite{Kouchi}. In their case any information of the molecular state has to be extracted from the pair of photons produced in the decay of the atomic fragments. In this case, no manipulation of the atoms themselves is possible due to the short life time of the handled atoms.

This work is supported by FAPERJ, CNPq and the Scientific Cooperation Agreement CAPES/COFECUB between France and Brazil, project number Ph 636/09.
F. Z. gratefully acknowledge FAPEMIG for financial  support.


\begin{thebibliography}{11}

\bibitem{Leventhal67} M. Leventhal, R. T. Robiscoe, and K. R. Lea, Phys. Rev. {\bf 158}, 49--56 (1967).

\bibitem{alkhalili04}A. Al-Khalili et al., {\em J. Chem. Phys.} {\bf{121}}, 5700 (2004).

\bibitem{wolf05} D Zajfman, A Wolf, D Schwalm, D A Orlov, M Grieser, R von Hahn, C P Welsch, J R Crespo Lopez-Urrutia, C D Schröter, X Urbain and J Ullrich,   J. Phys.: Conf. Ser. {\bf 4}, 296 (2005).

\bibitem{zhaunerchyk07}V. Zhaunerchyk, A. Al-Khalili, R.D. Thomas, W.D. Geppert, V. Bednarska, A. Petrignani, A. Ehlerding, M. Hamberg, M. Larsson, S. Ros\'{e}n, and W.J. van der Zande, \emph{Phys. Rev. Lett.} {\bf 99}, 013201 (2007). 

\bibitem{zande09}W. v.d. Zande (editor), {\em J. Phys.:Conference. Series} {\bf 192} (2009).

\bibitem{larsson-orel08}M. Larsson and A. E. Orel, "Dissociative Recombination of Molecular Ions", Cambrindge University Press (2008) 

\bibitem{Worner07} H.J. W\"{o}rner, S. Mollet, Ch. Jungen, and F. Merkt, Phys. Rev. A \textbf{75}, 062511 (2007).

\bibitem{Melero06} E. Melero Garc\'\i a, J. \'{A}lvarez Ruiz, S. Menmuir, E. Rachlew, P. Erman, A. Kivim\"{a}ki, M. Glass-Maujean, R. Richter, and M. Coreno, J. Phys B: At. Mol. Opt. Phys \textbf{39}, 205 (2006).

\bibitem{Bozek06} J. D. Bozek, J. E. Furst, T. J. Gay, H. Gould, A. L. D. Kilcoyne, J. R. Machacek, F. Mart\'\i n, K. W. McLaughlin, and J. L. Sanz-Vicario, J. Phys. B: At. Mol. Opt. Phys. \textbf{39}, 4871 (2006).

\bibitem{Sanchez97.99} I. S\'{a}nchez and F. Mart\'\i n, J. Chem. Phys. \textbf{106}, 7720 (1997); J. Chem. Phys. \textbf{110}, 6702 (1999).


\bibitem{Vanne06} Y. V. Vanne, A. Saenz, A. Dalgarno, R. C. Forrey, P. Froelich, and S. Jonsell, Phys. Rev. A \textbf{73}, 062706 (2006).

\bibitem{Martin07} F. Mart\'\i n, J. Fern\'{a}ndez, T. Havermeier, L. Foucar, Th. Weber, K. Kreidi, M. Sch\"{o}ffler, L. Schmidt, T. Jahnke, O. Jagutzki, A. Czasch, E. P. Benis, T. Osipov, A. L. Landers, A. Belkacem, M. H. Prior, H. Schmidt-B\"{o}cking, C. L. Cocke, R. D\"{o}rner, Science \textbf{315}, 629 (2007).


\bibitem{mot08} O. Motapon, Franois Olivier Waffeu Tamo, X. Urbain and I. F. Schneider, \emph{Phys. Rev.} \textbf{A77}, 052711 (2008).

\bibitem{Medina11} Aline Medina, G. Rahmat, C. R. de Carvalho, Ginette Jalbert, F. Zappa, R. F. Nascimento, R. Cireasa, N. Vanhaecke, Ioan F. Schneider, N. V. de Castro Faria and J. Robert, J. Phys. B: At. Mol. Opt. Phys. \textbf{44}, 215203 (2011).

\bibitem{Medina12} A. Medina, G. Rahmat, G. Jalbert, R. Cireasa, F. Zappa,  C. R. de Carvalho, N. V. de Castro Faria and J. Robert, Eur. Phys. J. D \textbf{66},  134 (2012).

\bibitem{Beguin} E.A. Power and T. Thirunamachandran, Phys.Rev. A \textbf{51}, 3660 (1995); L. B\'eguin, A. Vernier, R. Chicireanu, T. Lahaye, and A. Browaeys,  Phys. Rev. Lett. \textbf{110}, 263201 (2013).

\bibitem{Bohm} D. Bohm \textit{Quantum Theory} (New Jersey: Eglewood Cliffs/ Prentice-Hall, (1951).

\bibitem{Einstein} A. Einstein, B. Podolsky and N. Rosen, Phys. Rev. \textbf{47}, 777--780 (1935).

\bibitem{Schwinger} B-G. Englert, J. Schwinger and M. O. Scully, Found. Phys. \textbf{18},1045--1056 (1988). 

\bibitem{Fry} E. S. Fry, T. Walther and S. Li, Phys. Rev. A \textbf{52},4381-4395 (1995); J. Koperski and E. S. Fry, J. Phys. B: At. Mol. Opt. Phys. \textbf{39}, S1125 (2006); J. Koperski, M. Strojecki, M. Krosnicki, and T. Urbanczyk J. Phys. Chem. A \textbf{115}, 6851 (2011).

\bibitem{Kouchi} T. Tanabe, T. Odagiri, M. Nakano, I. H. Suzuki and N. Kouchi, Phys. Rev. Lett \textbf{103}, 173002 (2009).





\end{thebibliography}
\end{document}